\documentclass[conference]{IEEEtran}

\usepackage[T1]{fontenc}

\usepackage{amssymb}	 
\usepackage{amsmath}

\usepackage{graphicx}
\usepackage{booktabs}
\usepackage{multirow}
\usepackage{tabularx}
\usepackage{enumitem}
\usepackage{threeparttable}

\usepackage{balance}

\usepackage{url}

\binoppenalty=\maxdimen
\relpenalty=\maxdimen

\begin{document}

\title{Whose Hands Are in the Finnish Cookie Jar?}
\author{
\IEEEauthorblockN{Jukka Ruohonen}
\IEEEauthorblockA{Department of Future Technologies \\
University of Turku, Finland \\ 
Email: juanruo@utu.fi}
\and
\IEEEauthorblockN{Ville Lepp\"anen}
\IEEEauthorblockA{Department of Future Technologies \\
University of Turku, Finland \\ 
Email: ville.leppanen@utu.fi}
}

\maketitle

\begin{abstract}
Web cookies are ubiquitously used to track and profile the behavior of users. Although there is a solid empirical foundation for understanding the use of cookies in the global world wide web, thus far, limited attention has been devoted for country-specific and company-level analysis of cookies. To patch this limitation in the literature, this paper investigates persistent third-party cookies used in the Finnish web. The exploratory results reveal some similarities and interesting differences between the Finnish and the global web---in particular, popular Finnish web sites are mostly owned by media companies, which have established their distinct partnerships with online advertisement companies. The results reported can be also reflected against current and future privacy regulation in the European Union.
\end{abstract}


\section{Introduction}

%

This short empirical paper surveys the use of persistent cookies in the contemporary Finnish world wide web. The term Finnish web is understood as a sub-population of popular web sites that are primarily designed for Finnish users, using Finnish as the primary language for the content. This previously unused country-specific focus establishes the paper's contribution to the existing empirical cookie research. Another contribution is made
with a company-level analysis, which reveals interesting patterns in the relationships between Finnish media companies and global online advertisement companies.

\section{Exploratory Analysis}

\subsection{Data}\label{subsec: data}

The sample contains $10,240$ persistent cookies that were collected from $206$ popular Finnish web sites. The sampling from a little over two hundred sites is on the small side; prior studies have collected cookies from a ten or even a hundred thousand web sites \cite{Aleyasen15, Cahn16a, Tappenden09}. While such numbers are necessary when attempting to infer about the global world wide web, the number of web sites sampled is reasonable because the Finnish web is relatively small and highly concentrated in terms of popularity. Current estimates from early 2017 indicate that the two most popular sites (\texttt{is.fi} and \texttt{yle.fi}) both had a weekly reach of about two million Finnish people~\cite{TNS17}, which is nearly forty percent of the whole population in Finland.  Given the high concentration, it is a reasonable assumption that sampling from the $206$ most popular web sites generalize well toward the statistical ``population'' of Finnish people who visit Finnish web sites at least once a week.

The web sites used for the cookie harvesting were collected from the weekly popularity statistics maintained by the Finnish market research company Kantar TNS Oy \cite{TNS17}. Although the methodological details are proprietary, the company uses cookie-based tracking, traffic analysis, and other means to rank the popularity of Finnish sites. Unlike Alexa Internet, whose popularity lists have been used in previous cookie research~\text{\cite{Cahn16a, Tappenden09, Acar14, Olejnik14}}, TNS uses also survey data from Finnish web users, which presumably further increases the robustness of the popularity ranks. Given these supposedly rather reliable ranks, the sampled web sites were collected by merging all unique sites that were ranked between 2000 and 2017, using the first week of January as the annual reference.

The persistent cookies from the popular web sites sampled were collected in 14 April 2017 with Firefox version 45.8.0, using the browser's default settings. All cookies were retrieved from Firefox's \texttt{cookies.sqlite} file, and after each visit, all Firefox-specific data was deleted from a local hard disk. Finally, a ten second waiting time was used between visits in order to ensure that full contents were successfully retrieved.

\subsection{First and Third Parties}

Cookies can be divided into session cookies and persistent cookies. Due to the limitations in the hypertext transfer protocol (HTTP), session cookies are generally required for most current web applications for maintaining state information. Such cookies are erased when a browser is closed, whereas persistent cookies are retained across multiple browser sessions. All cookies observed in this paper are persistent.

Persistent cookies allow storing users' preference of a web site, but these are frequently used also for tracking and profiling of users. The further distinction between first-party and third-party cookies reflects this dual functionality implicitly. According to the same origin policy used by web browsers, an HTTP request made by a browser contains only those cookies whose domain attribute matches the domain name of the request (see \cite{Rabinovich3} and \cite{Kristol01} for summaries of the technical and historical background). When querying \texttt{hs.fi}, for instance, the cookies set for the domains \texttt{adtech.de} and \texttt{hs.fi} are third-party and first-party cookies, respectively. The cross-domain sharing of information between the two domains can occur in multiple ways, but in a typical scenario \texttt{hs.fi} would contain advertisements served from the domain \texttt{adtech.de}.

\begin{figure}[th!b]
\centering
\includegraphics[width=7cm, height=5cm]{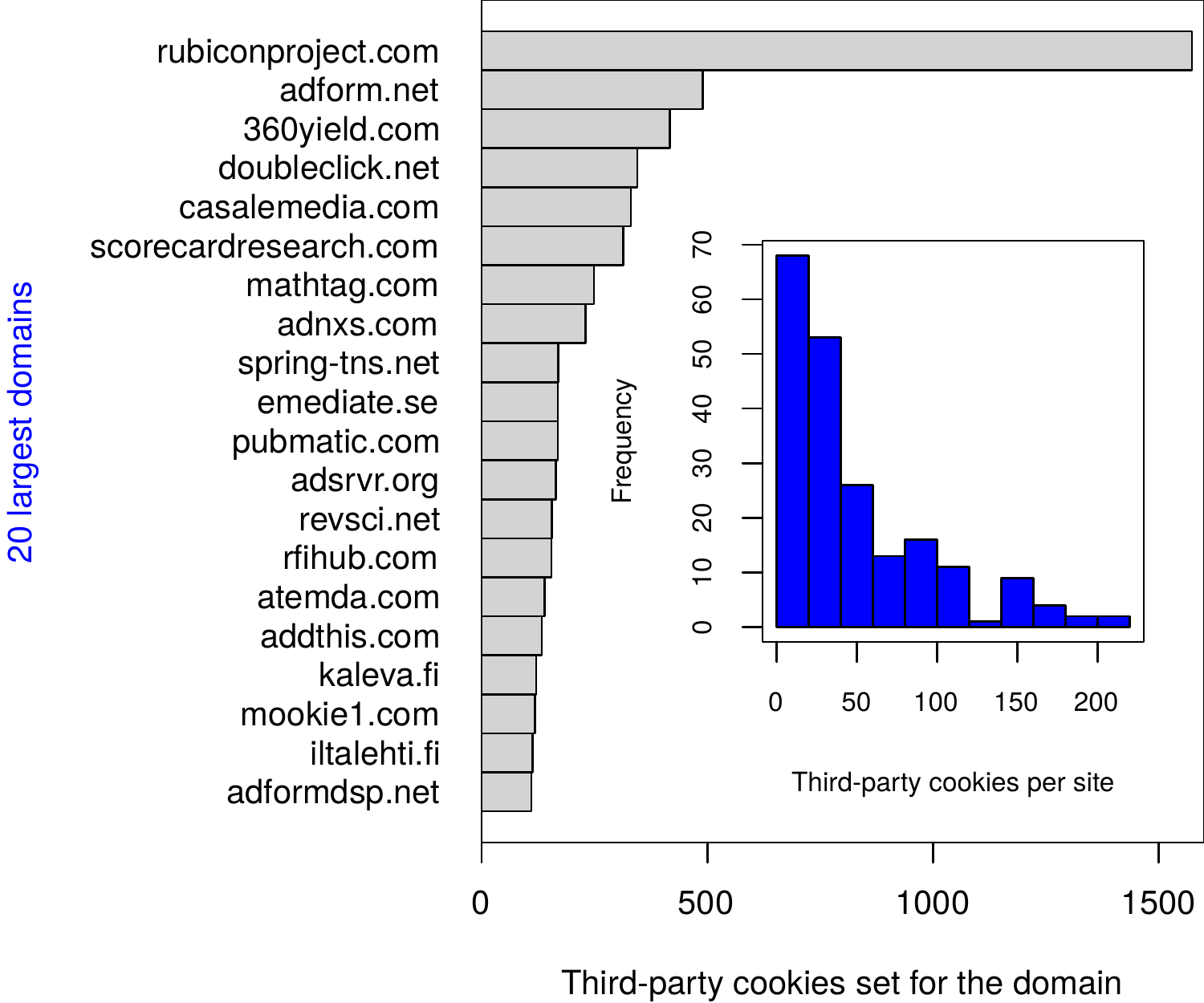}
\caption{A Summary of Third-Party Cookies Used in the Finnish Web}
\label{fig: third-party}
\end{figure}

Given this background, it is telling to start the exploration with an observation that about 91.4~\% of cookies in the sample are third-party. The rate is higher than what has been observed in the global web \cite{Cahn16a, Tappenden09}. As an outlook on the actual third-party domains, Fig.~\ref{fig: third-party} summarizes the amount of third-party cookies set for particular domains (outer plot) and the distribution of third-party cookies per web site (inner plot). To further examine the hypothesis that increasing popularity would entail heavy use of cookies \cite{Tappenden09}, Table~\ref{tab: correlations} shows also correlation coefficients between the number of third-party cookies and five popularity metrics made available by TNS~\cite{TNS17}. 

\begin{table}[th!b]
\centering
\begin{threeparttable}
\caption{Correlations Between the Number of Third-Party Cookies and Five Popularity Metrics (Pearson $r$)}
\label{tab: correlations}
\begin{tabular}{lccccc}
\toprule
& Rank & Users & Browsers & Sessions & Page views \\
\cmidrule{2-6}
$r$ & $-0.02$ & $0.09$ & $0.07$ & $0.09$ & $0.09$ \\
\bottomrule
\end{tabular}
\begin{scriptsize}
\begin{tablenotes}
\item[]{\textit{Nota bene}: all metric values are averages from the eight weekly snapshots used for the sampling (see Section~\ref{subsec: data}). The five popularity metrics are: TNS's popularity rank (akin to Alexa); average weekly reach among 15--74 year old Finnish people; and the average weekly number of browsers, sessions, and page views, in the order of listing.}
\end{tablenotes}
\end{scriptsize}
\end{threeparttable}
\end{table}

Three observations can be made from the results. First and foremost, the use of third-party cookies is extensive but varies across sites. While most Finnish sites use less than $50$ third-party cookies, there are sites storing even up to $200$ persistent third-party cookies. These numbers alone are enough to raise some privacy concerns. Second, the most frequent tracking domains show some divergence from the global web~(cf.~\text{\cite{Cahn16a, Olejnik14, Cahn16b}}). For instance, Rubiconproject is the clear leader, while Google's \texttt{doubleclick.net} takes only the fourth place. Last, the correlation coefficients reported are all modest: the popularity of Finnish web sites neither increases nor decreases the amount of third-party cookies.

\subsection{Media Companies and Their Tracking Partners}

Genre analysis has been one way to shed further light on persistent third-part cookies \cite{Cahn16a, Olejnik14, Gomer13}. With few e-commerce and social media platforms thrown in, the most popular Finnish web sites are online versions of newspapers, periodicals, and television channels. Almost all of these are either owned by media companies or, in the case of \texttt{yle.fi}, the state. Therefore, in the present context a slightly better classification can be done according to the owners of the  sites.


\begin{figure}[th!b]
\centering
\includegraphics[width=7cm, height=5cm]{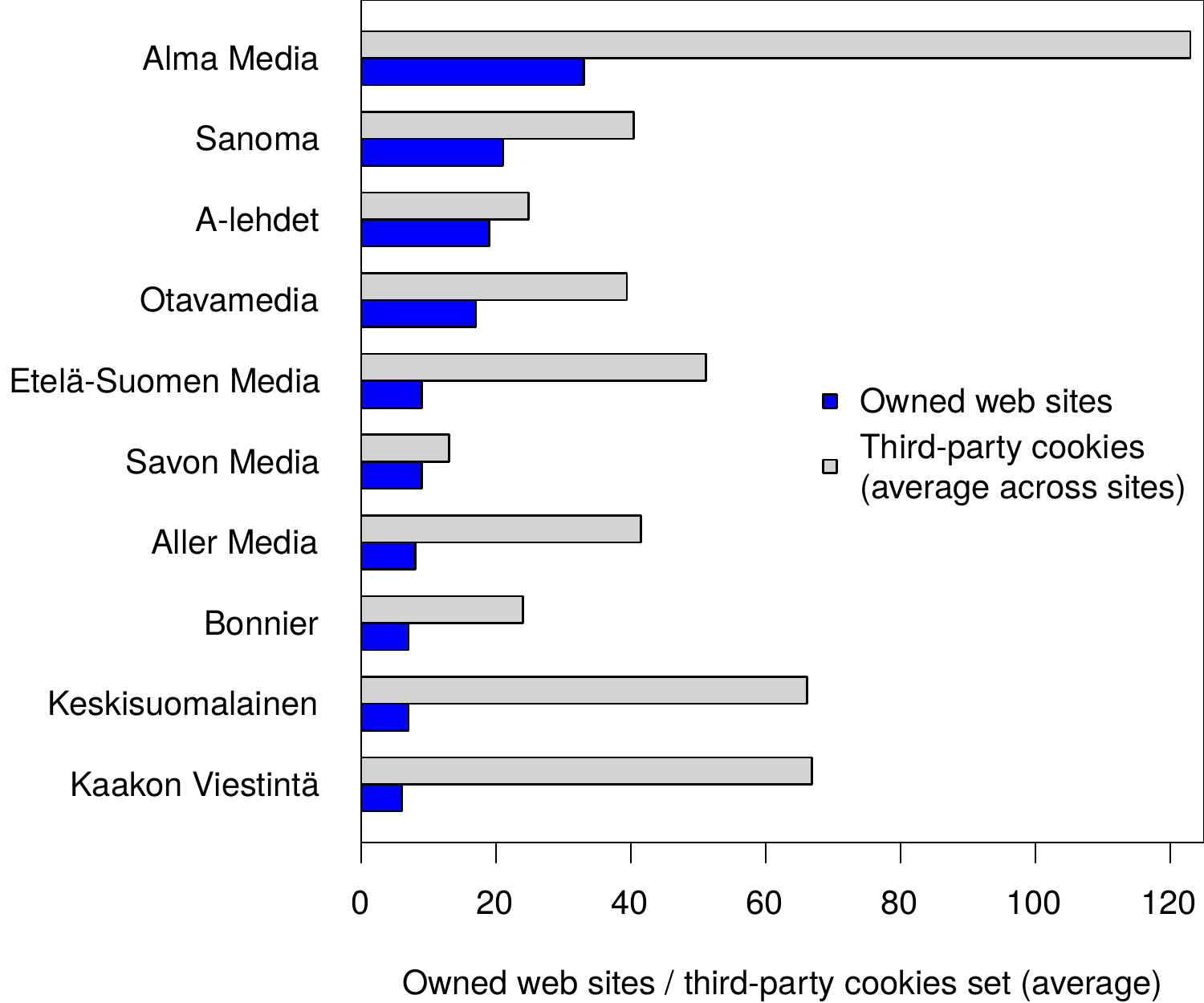}
\caption{The Most Popular Finnish Sites According to Top-10 Owners}
\label{fig: companies}
\end{figure}


\begin{figure}[th!b]
\centering
\includegraphics[width=7cm, height=5cm]{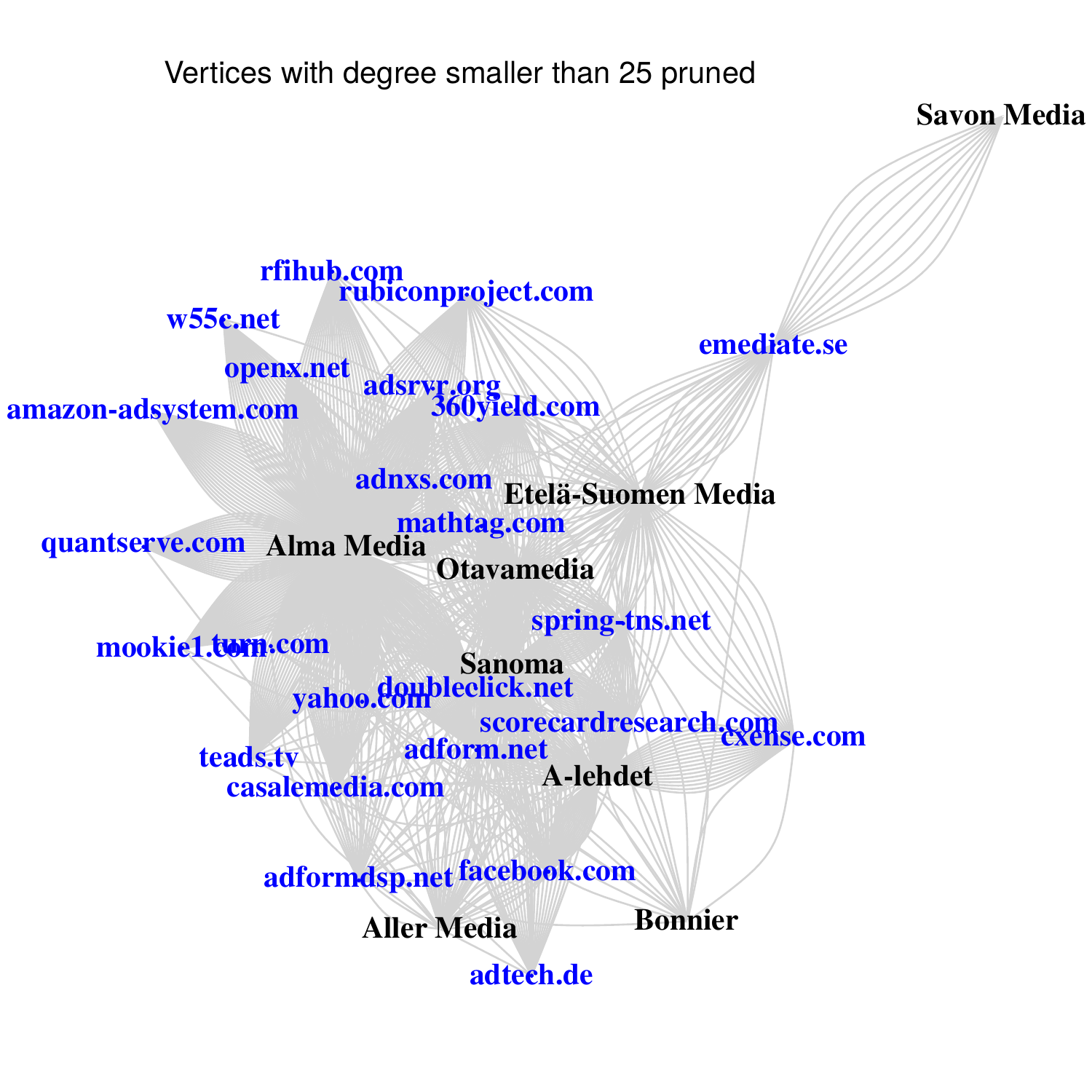}
\caption{Pruned Third-Party Cookie Network of Top-8 Web Site Owners}
\label{fig: tracking network 2}
\end{figure}

According to the results summarized in Fig.~\ref{fig: companies}, most of the web sites sampled are owned by publishing companies. Also the average number of third-party cookies set by the web sites vary according to the companies. In particular, the web sites owned by the publishing company Alma Media seem to set an extensive amount of third-party cookies. Some of these are used for sharing cross-domain information between the company's web sites (for instance, \texttt{talouselama.fi} stores a third-party cookie for \texttt{iltalehti.fi}, which is natural as both newspaper sites are owned by Alma Media). Many of the third-party cookies are set for known trackers, however. For visualizing the main trackers, Fig.~\ref{fig: tracking network 2} displays a bipartite network of the main tracking domains across eight companies. Interestingly, not all of the companies are using the same trackers; again, the web sites owned by Alma Media use a few trackers not used by the other media companies.

\subsection{Values}

The following listing raises a few illuminating points about the values stored to the persistent cookies collected.

\begin{itemize}
\item{As many as 86 web sites---about 40~\% of the web sites sampled---store the Internet protocol address used for the collecting in plain-text to third-party cookies.}
\item{Furthermore, two web sites store a private locate area network address (\texttt{10.218.140.241}) to a third-party cookie (\texttt{rubiconproject.com}) in plain text. This address may denote the server that was used for delivering the web content, but other scenarios are also possible.}
\item{Numerous sites store the visiting time in plain-text either as a timestamp or as a calendar time string. One site stores the user-agent string to a third-party cookie in plain text.}
\item{Five web sites store some fragments from a ``GeoIP'' identification to third-party cookies; the geographic location used for the queries (Helsinki) appears in plain text.}
\item{It is dreadful to end the enumeration with an observation that two popular sites store session and login information in plain text to first-party cookies. The stored information includes user name, user's first name, last name, company, and address, as well as session identifiers.}
\end{itemize}

These details indicate that quite a few tracking web sites do not bother to follow the basic guideline about hashing the values stored to cookies. Potential privacy leaks are also present. While addresses and geographic locations are leaked to third-party sites also via many other channels (such as advertisements served via \texttt{iframe}'s), some of the plain-text information stored to third-party cookies hint about further tracking efforts via browser fingerprinting (cf.~\cite{Nikiforakis15}). The last point in the listing raises also security concerns, which are also evident by further taking a look on the cookie attributes.



\subsection{Attributes}

The following brief observations can be made about the optional attributes set for the persistent cookies harvested.

\begin{itemize}
\item{If the so-called ``\textit{Secure}'' attribute is set for a cookie, the cookie should be transmitted only via secured (HTTPS) requests \cite{RFC6265}. Only 0.33~\% of the cookies collected have set this binary attribute. This negligible share is practically identical to the share (0.36~\%) recently observed in a sample collected from the global world wide web~\cite{Cahn16a}.}
\item{If the so-called ``\textit{HttpOnly}'' attribute is set for a cookie, the cookie should not be accessible through client-side (in-browser) scripts~\cite{RFC6265}. This optional attribute was originally introduced by Microsoft in the early 2000s for preventing a certain class of cross-site scripting (XSS) attacks \cite{OWASP17}. Alas, adoption has been slow: according to empirical surveys, the estimated global adoption rates range from 6.5~\%~\cite{Cahn16a} to 33.5~\%~\cite{vanGoethem2014}. The modest adoption rate applies also to the contemporary Finnish web: only about~7.5~\% of the cookies observed have set this flag.}
\item{The so-called ``\textit{path}'' attribute can be used to restrict the scope of a cookie to specific web server directories. When the path is either not specified or it is set to the web server root (\texttt{/}), a given cookie is submitted by a browser to all web applications accessible from the same domain name. Therefore, a basic risk relates to using unrestricted cookies for a deployment that  contains security-critical and other web applications in different directories \cite{Stuttard08}. Akin to the previous two attributes, however, about 99.4~\% of the cookies observed have a root-level scope. This share is again comparable to a rate (98.4~\%) seen in a cookie sample from the global web~\cite{Cahn16a}.}
\item{The ``\textit{expiration}'' attribute can be used to announce when a given cookie should be invalidated by a browser. When using the date of collecting as the other reference point, the lifetime of the cookies observed is relatively short, although the distribution visualized in Fig.~\ref{fig: lifetime} is skewed due to two sites announcing expiration dates in \texttt{31-12-9999}. This said, the median lifetime of third-party cookies is about six months, which is clearly above the one month cut-off point sometimes used for separating transient cookies from tracking cookies~\cite{Acar14}. 
\begin{figure}[th!b]
\centering
\includegraphics[width=\linewidth, height=4cm]{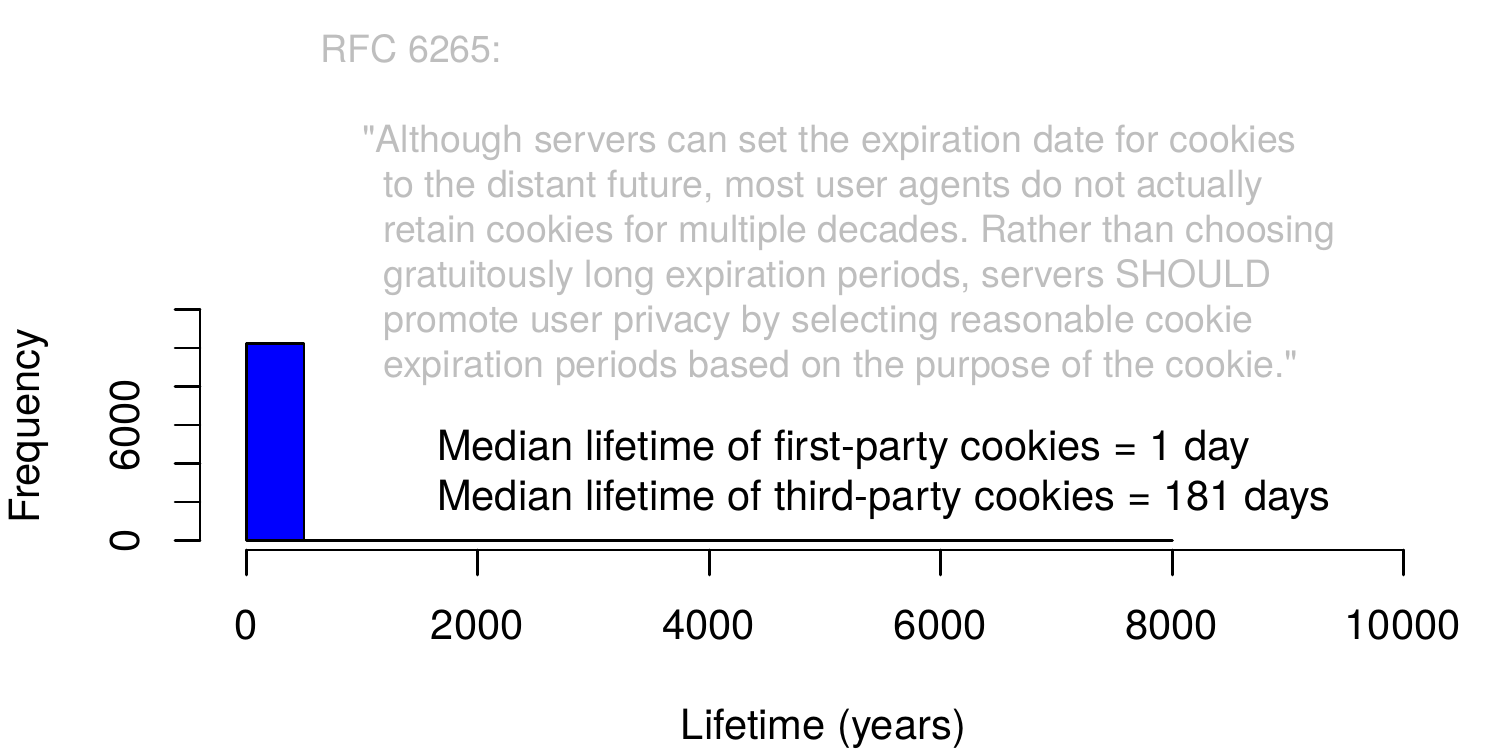}
\caption{The Cookie Lifetimes According to Expiry Dates}
\label{fig: lifetime}
\end{figure}
}
\end{itemize}

A good way to summarize the cookie attributes is the concept of maximal permissions, which are satisfied whenever cookies are persistent and non-secure with root-level scope~\cite{Cahn16b}. According to this definition, as much as 99.1~\% of the cookies observed have maximal permissions. It should be also remarked that the attributes outlined may be used primarily in cookies set during authentication to web applications~\text{\cite{Cahn16a, Calzavara14}}. Although such cases are not present in the sample, it is fair to tentatively conclude that cookie attributes are mostly ignored by the web sites sampled, which is unfortunate as these provide some security benefits for both servers and clients.

\subsection{Do Not Track}

\begin{figure}[th!b]
\centering
\includegraphics[width=\linewidth, height=4cm]{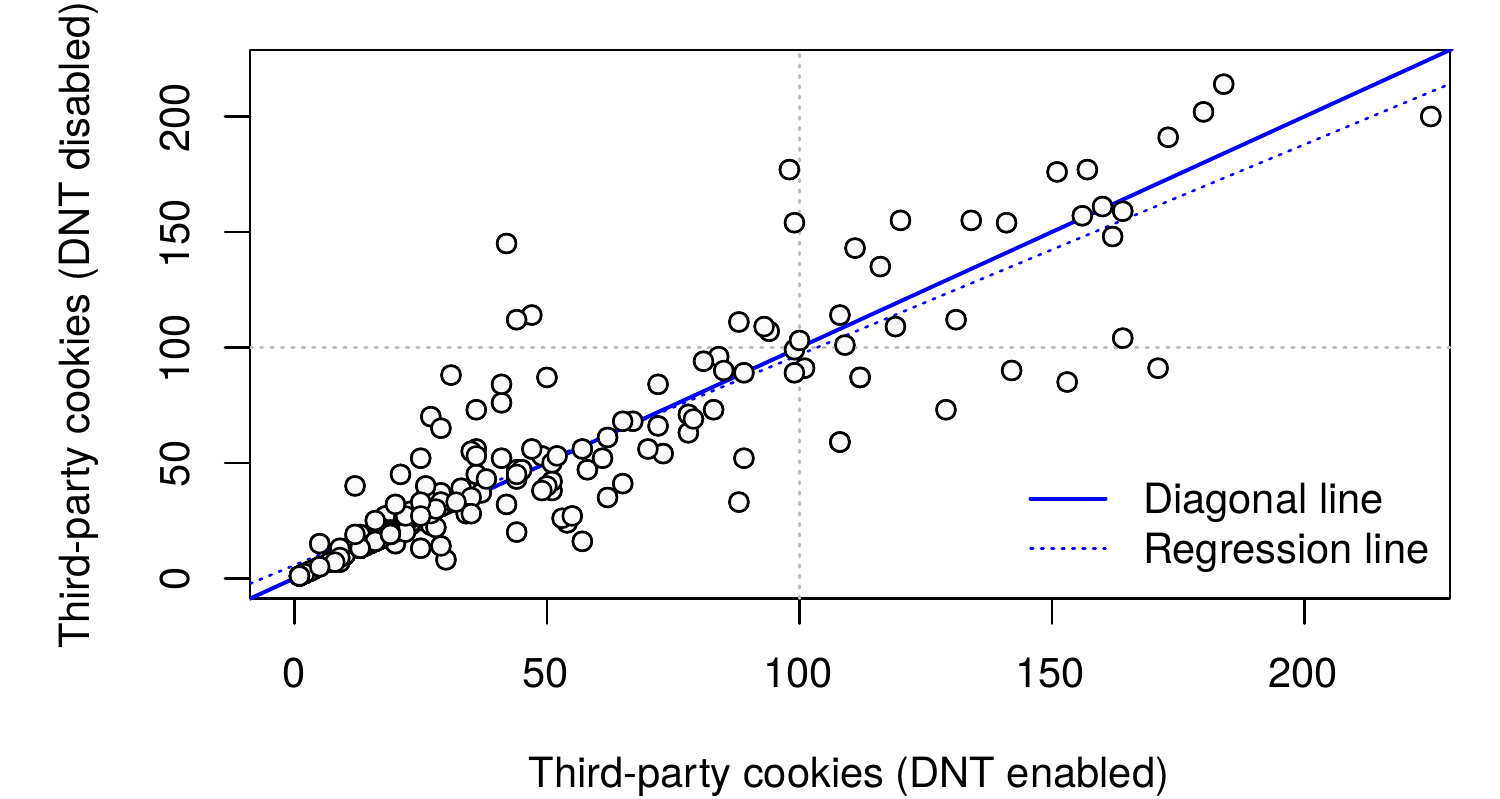}
\caption{Number of Third-Party Cookies with DNT Enabled and Disabled}
\label{fig: dnt}
\end{figure}

The so-called ``do not track'' (DNT) initiative was first proposed in 2007, and the controversy has endured ever since. Industrial adoption has been slow \cite{Acar14}. Although DNT does not block third-party cookies as such, it is interesting to briefly examine whether the number of third-party cookies vary according to whether DNT is enabled or disabled. The results visualized in Fig.~\ref{fig: dnt} are based on repeating the cookie collection routine (see Section~\ref{subsec: data}) with DNT enabled in Firefox (the $y$-axis and $x$-axis display the number of third-party cookies with DNT disabled and enabled, respectively).

To read the figure, values above the diagonal line are ``good for privacy'' in the sense that enabling DNT resulted fewer third-party cookies. While some of the web sites indeed store fewer third-party cookies with DNT, many sites are also located below the diagonal, meaning that actually more third-party cookies were stored upon enabling DNT in Firefox. As a concrete example: when DNT is announced by a browser, some web sites store a value \texttt{``DNT2''} to third-party cookies set for the domain \texttt{exelator.com}. All in all, however, the regression line shown is close to the diagonal line, meaning that DNT does not entail a noteworthy difference.

\section{Discussion}

This short empirical paper explored the use of cookies in the Finnish web. Like everywhere, (a) the use of third-party tracking cookies is rampant in the Finnish web. According to the results, it also seems that (b) the Finnish web does not differ notably from the global web in terms of cookie attributes, but (c) there are some eminent differences in terms of the online advertisement companies used by the owners of popular Finnish web sites. It is also evident that (d) some of these owners prefer partnerships with smaller advertisement companies instead of relying on the global market leaders. Finally, (e) web site popularity and the ``do not track'' initiative do not statistically explain the amount of third-party cookies.

Almost all of the Finnish web sites sampled display a notification about the use of cookies, following the regulation imposed by the European Union (EU) regarding user consent for the use of cookies (see \cite{Leenes15} and \cite{Legge15} for the legal background). Many Finnish web site also point toward the European industry consortium that allows users to opt-out from tracking---by ironically installing a further tracking cookie \cite{EDAA}. As such opt-out solutions are dubious in many respects~\cite{Leon12}, further legislative amendments may be required as an alternative to industry self-regulation. For instance: if the EU privacy regulation continues to require consent from users, it might be worthwhile to also mandate web site owners to explicitly enumerate all domains stored to third-party cookies. Another question is how well the Finnish web sites and their users truly understand the security and privacy consequences from serving content from third-party domains.

\section*{Acknowledgements}

The authors gratefully acknowledge Tekes -- the Finnish Funding Agency for Innovation, DIMECC Oy, and the Cyber Trust research program for their support.

\balance


\end{document}